\documentclass[a4paper]{article}

\usepackage{INTERSPEECH2022}
\usepackage{todonotes}
\usepackage{enumitem}

\title{Predicting within and across language phoneme recognition performance of self-supervised learning speech pre-trained models}
\name{Hang Ji, Tanvina Patel, Odette Scharenborg}
\address{
  Multimedia Computing Group, Delft University of Technology, Delft, The Netherlands}
\email{H.Ji@student.tudelft.nl, T.B.Patel@tudelft.nl, O.E.Scharenborg@tudelft.nl}

\begin{document}

\maketitle
\begin{abstract}

In this work, we analyzed and compared speech representations extracted from different frozen self-supervised learning (SSL) speech pre-trained models on their ability to capture articulatory features (AF) information and their subsequent prediction of phone recognition performance for within and across language scenarios. Specifically, we compared CPC, wav2vec 2.0, and HuBert. First, frame-level AF probing tasks were implemented. Subsequently, phone-level end-to-end ASR systems for phoneme recognition tasks were implemented, and the performance on the frame-level AF probing task and the phone accuracy were correlated. Compared to the conventional speech representation MFCC, all SSL pre-trained speech representations captured more AF information, and achieved better phoneme recognition performance within and across languages, with HuBert performing best. The frame-level AF probing task is a good predictor of phoneme recognition performance, showing the importance of capturing AF information in the speech representations. Compared with MFCC, in the within-language scenario, the performance of these SSL speech pre-trained models on AF probing tasks achieved a maximum relative increase of 34.4\%, and it resulted in the lowest PER of 10.2\%. In the cross-language scenario, the maximum relative increase of 26.7\% also resulted in the lowest PER of 23.0\%.


\end{abstract}

\section{Introduction}




Recently, the self-supervised learning (SSL) speech pre-trained models have shown great potential in the field of speech technology \cite{yang2021superb}. These SSL speech pre-trained models are trained on huge amounts of unlabeled raw speech using self-supervised learning, and learn speech representations which are conditional dependent on different parts of the input raw speech sequence \cite{oord2018representation, baevski2020wav2vec, hsu2021hubert}. They can be used directly  \cite{nguyen2020zero,blandon2020analysis}, or can be fine-tuned in a specific speech downstream task \cite{baevski2019effectiveness} to achieve great improvements.
Several SSL speech pre-trained models have recently been proposed. For example, the contrastive predictive coding (CPC) speech pre-trained model, which is trained by minimizing the contrastive predictive loss of predicted frames \cite{oord2018representation}, and wav2vec 2.0 \cite{baevski2020wav2vec} 
which uses the transformer architecture to generate speech representations which are dependent on the entire preceding input sequence up to the one under consideration. SSL speech pre-trained models are assumed to discard information that is not important for speech recognition, such as noise and speaker information, while retain acoustic information that is important for improving speech recognition  \cite{oord2018representation,baevski2020wav2vec,hsu2021hubert,chung2019unsupervised}. 

With the development of these SSL speech pre-trained models, several papers have compared different models to investigate their intelligibility and potential. For instance, Yang et al. \cite{yang2021superb} introduced a benchmark for comparing the performance of different SSL speech pre-trained models on different speech downstream tasks including automatic speech recognition (ASR), keyword spotting, and speech enhancement, etc. Their work showed SSL speech pre-trained models perform better than a widely-used feature, FBANK \cite{yang2021superb}, with a large margin. Pasad et al. \cite{pasad2021layer} implemented a suite of analysis tools based on non-parametric probes to reveal different information encoded in intermediate representations of a speech pre-trained model, wav2vec 2.0. Moreover, speech pre-trained models transfer well across languages \cite{riviere2020unsupervised, babu2021xls}. For example, the speech pre-trained model CPC, which is trained on English, could also perform well in Dutch phoneme recognition \cite{riviere2020unsupervised}. In addition, the transferability of a speech pre-trained model across different languages was analyzed by a distance-based metric, ABX score, on discovered phonemes. The distance-based metric indicates how phones are separated into different phonemes by speech representations generated by the SSL speech pre-trained model.

An open question that is addressed in this work, is what linguistic information is encoded in the speech representations learned by these SSL speech pre-trained models and how this relates to a downstream task. This paper aims to quantitatively assess the phonetic information captured by three state-of-the-art SSL speech pre-trained models: CPC, wav2vec 2.0, and HuBert. Specifically, we investigate what articulatory feature information is encoded in the SSL speech representations extracted by the  different speech pre-trained models, and how this relates to the performance on a downstream phoneme recognition task. To that end, a frame-level articulatory feature (AF) probing task is used to analyze the articulatory information encoded in the speech representations. Subsequently, the same SSL speech pre-trained models are used as the feature extractor to extract feature representations of an phone-level end-to-end  ASR system, after which the results of the AF probing task and the phoneme recognition task are correlated to answer the following two research questions: 
\vspace{-1mm}
\begin{itemize}[leftmargin=*]
\item \textbf{RQ1, within-language:} What articulatory feature (AF) information is modelled by the SSL speech pre-trained models and how does this correlate to phoneme recognition performance in the same language, i.e., English?
\item \textbf{RQ2, cross-language:} To what extent is the AF information from a different language modelled by the SSL speech pre-trained models and how does it correlate to phoneme recognition performance in the other language, specifically Mboshi, an African Bantu language? 
\end{itemize}
\section{Methodology}
\begin{figure}
  \centering
  \includegraphics[width=.6\columnwidth]{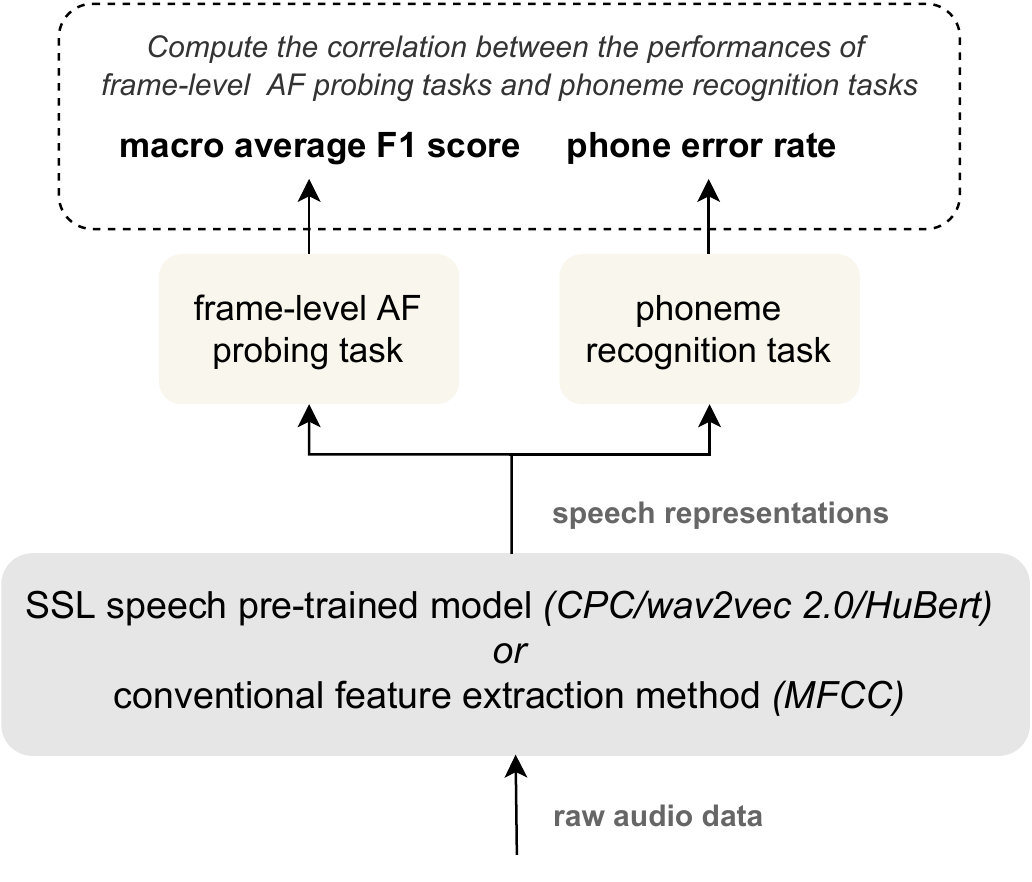}
  \caption{Overview of the experimental set-up of this work.}
  \label{fig:method}
\vspace{-4mm}
\end{figure}

Figure \ref{fig:method} gives an overview of our experimental set-up. The three SSL speech pre-trained methods  (Section 2.1), Contrastive Predictive Coding, wav2vec 2.0, and HuBert, provide 1) input to frame-level articulatory feature probing tasks to analyze and compare the AF information encoded in their speech representations (Section 2.2). 2) Subsequently, these SSL speech pre-trained models are used as the feature extractor in phoneme recognition tasks (Section 2.3). As a baseline, the same experiments will be carried out while using MFCCs as input. Finally, results of the AF probing task and the phoneme recognition task are strongly correlated\footnote{Implementation: https://github.com/KarenMars/IS22Code}.
\subsection{Self-supervised learning speech pre-trained models}

\textbf{Contrastive Predictive Coding (CPC) \cite{oord2018representation}}: The architecture includes an encoder module and a context module. The encoder module generates a latent representation \(z_t=g_{enc}(x_t)\) from a sequence of raw audio data \(x_t\). Afterwards, the context module generates the context representation \(c_t=g_{con}(z_{t})\) which is conditionally dependent on the context representations of previous time steps. Typically, the encoder consists of a multi-layer convolutional neural network, and the context module consists of recurrent neural networks. The training objective is to minimize the contrastive loss, which aims to maximize the mutual information between the future latent representation \(z_{t+k}\) in \(k\) time steps ahead and the prediction value \(W_kc_t\), and minimize the mutual information between the latent representation \(z_{j}\) from negative samples \(Z = \{z_1,...z_N\}\) and the prediction \(W_kc_t\). \(W_k\) performs linear transformation of \(c_t\). The training objective is optimized by minimizing the sum of the loss for different time step \(k\), \(k \in \{1, ...K\}\), following:
\begin{equation}
    \mathcal{L}_{CPC}=-\frac{1}{K}\sum_{k=1}^K\left[log\frac{exp(z_{t+k}^TW_kc_t)}{\sum_{{z_{j}}\in Z}exp(z_{j}^TW_kc_t)}\right]
\end{equation}
\noindent \textbf{wav2vec 2.0 \cite{baevski2020wav2vec}}: Similar to CPC, wav2vec 2.0 also has an encoder module \(z_t=g_{enc}(x_t)\) and an context module \(c_t=g_{con}(z_{t})\). Additionally, it contains a quantization module, which converts the continuous latent representations \(z_t\) to a discrete representation \(q_t=g_{quan}(z_t)\). The encoder consists of a multi-layer convolutional neural network and the context module consists of a Transformer network. The quantization module utilizes product quantization \cite{jegou2010product}, which converts latent representations to discrete representations by concatenating entries sampled from different codebooks. Similar to CPC, the training objective of wav2vec2.0 is to minimize the contrastive loss. Specifically, a portion latent representations \(z_t\) generated by the encoder module from input time steps are masked before being fed to the context module. Afterwards, wav2vec2.0 aims to identify the true quantized latent speech representation \(q_t\) instead of the true latent speech representation as in CPC for the input of a masked time step. It is also known as the masked prediction loss, \(\mathcal{L}_m\). Moreover, the training objective is augmented by a code diversity loss \(\mathcal{L}_d\), which ensures the equal use of codebook entries from different codebooks. The training objective is as follows:
\begin{equation}
    \mathcal{L}_{wav2vec 2.0}=\mathcal{L}_m+ \alpha\mathcal{L}_d
\end{equation}
\begin{equation} \label{eq: masked prediction loss}
    \mathcal{L}_m=-log\frac{exp(sim(c_t,q_t))}{\sum_{{\tilde{q}\in Q_t}}exp(sim(c_t,\tilde{q}))}
\end{equation}
where \(\alpha\) is a tuned hyperparameter, \(Q_t\) is the set of quantized candidate representations, which consist of the true sample and negative samples from other masked time steps.

\noindent \textbf{HuBert \cite{hsu2021hubert}}: The implementation of HuBert, which also consists of an encoder \(z_t=g_{enc}(x_t)\) and a context module \(c_t=g_{con}(z_{t})\), is identical to wav2vec 2.0. However, instead of the quantization module used in wav2vec 2.0, HuBert utilizes an offline acoustic unit discovery (AUD) module, which is a clustering module such as k-means. Before training HuBert, the AUD module assigns the related cluster to each frame \(x_t\) of the input raw audio data as its pseudo label \(u_t \in [C]\). Similar to wav2vec 2.0, HuBert adopts the contrastive loss. Specifically, HuBert aims to identify the true embedding  \(e_c\) of the pseudo label instead of the quantized latent speech representations. The unmasked time steps are also included in computing the contrastive loss. The training objective is as follows:
\begin{equation}
    \mathcal{L}_{HuBert}=\alpha\mathcal{L}_m+ (1-\alpha)\mathcal{L}_u
\end{equation}
where \(\alpha\) is a tuned hyperparameter, and  \(\alpha \in [0,1]\). \(\mathcal{L}_m\) or \(\mathcal{L}_u\) is similar to Equation \ref{eq: masked prediction loss}, with \(q_t\) replaced by the embedding \(e_c\) of the pseudo label \(u_t \in [C]\).

\subsection{Frame-level articulatory feature probing task}

The frame-level AF probing task is to analyze how well the speech representations encode different AF information, which is implemented as an AF classification task. AFs are acoustic correlates of how phones are produced by the vocal tract  \cite{Jurafsky:2009:SLP:1214993}. We use seven articulatory features in the probing task, and these features and their quantized classes are presented in Table \ref{tab:af}. The input of the AF classification task is the speech representation of a frame from the raw audio data. The output consists of the predicted classes for the seven different articulatory features. 
\begin{table}[ht]
\centering
\caption{Articulatory features and their quantized classes \cite{scharenborg2007towards}}
\label{tab:af}
\resizebox{\columnwidth}{!}{%
\begin{tabular}{cc}
\toprule
AF         & Values                                                              \\ \midrule
`manner'   & approximate, retroflex, fricative, nasal, stop, vowel, nil \\
`place'    & bilabial, labiodental, dental, alveolar, velar, nil        \\
`voice'    & +voice, -voice                                             \\
`high-low' & high, mid, low, nil                                      \\
`fr-back'  & front, central, back, nil                                 \\
`round'    & +round, -round, nil                                        \\
`static'   & static, dynamic                                            \\ \bottomrule
\end{tabular}%
}
\vspace{-4mm}
\end{table}
We use a multi-class support vector machine (SVM) for the AF classification task, one for each articulatory feature. Since the probing task is to reveal the encoded articulatory information of speech representations, a high nonlinear separability is not required for these classifiers. Thus, linear SVMs with soft-margin rather than SVMs with nonlinear kernels are adopted in the probing task. In addition, a one-versus-the-rest multiclass strategy \cite{bishop2006pattern} is used for the AFs with more than two AF classes: i.e., for each AF class of an AF, a binary classifier is used to discriminate the particular AF class from the other AF classes. For example, for the AF `round', three binary classifiers are trained for the three AF classes in order to discriminate one AF class from the others. At test, confidence scores for each binary classifier are computed for the tested frame, and the label of the class with the highest confidence score is assigned to the tested instance. The frame-level AF probing task is evaluated using the averaged macro-averaged F1 score, which is computed by the mean of all per-AF macro-averaged F1 scores. The macro-averaged F1 score is as follows: \(macro\textnormal{-}averaged~F1~score = \frac{1}{|C|}\sum_{i\in C}F1_i\), where \(C\) is the set of classes in a specific AF.
\subsection{Phoneme recognition system}

We use the hybrid CTC/attention end-to-end (E2E) ASR system \cite{watanabe2017hybrid} as our phoneme recognition system. The E2E consists of an encoder which is implemented by one layer of  Gated Recurrent Units (GRU), and a decoder which is implemented by one layer of LSTM. The system is trained with a joint CTC and attention objective function, as follows:
\begin{equation}
    \mathcal{L}=\lambda\log p_{ctc}(C|X) + (1-\lambda)\log p^{*}_{att}(C|X)
\end{equation}
where \(\lambda\) is a tuned hyperparameter, and \(\lambda \in [0,1]\), \(X\) is the sequence of speech representations, \(C\) is the output phoneme sequence given \(X\).
\section{Experimental details}
\subsection{Corpora}
We investigated the AF information and the performance of the speech representations on the phoneme recognition task on two languages: English, which is the same language as the pre-trained models were trained on, and Mboshi, a Bantu language, unrelated to English.

We trained the CPC model with LibriSpeech \cite{panayotov2015librispeech}, the database that was used to train the wav2vec 2.0  and HuBert models. LibriSpeech is a read English speech corpus including 2338 speakers. The probing and phoneme recognition tasks were carried out on TIMIT \cite{garofolo1993darpa} for RQ1 and the Mboshi database \cite{DBLP:journals/corr/abs-1710-03501} for RQ2. TIMIT is a read English speech corpus which is 5.4 hours long with 6300 sentences and read by 630 speakers. The number of phonemes in TIMIT is 39. And the Mboshi database is a read Mboshi speech corpus which is 4.9 hours long with 5130 sentences and read by 3 speakers. The number of phonemes in Mboshi is 68. Meanwhile, TIMIT and the Mboshi database contain the force-aligned phoneme transcriptions.

For probing tasks, the train-test split follows the split provided by the corpora, for TIMIT, 3696 sentences are used in the training data set, and 1344 sentences are used in the testing data set, and for Mboshi, the split is 4616 sentences and 514 sentences. For phone recognition tasks, for TIMIT, 3696 sentences are used in the training data set, 400 sentences are used in the validation data set, and 192 sentences are used in the test data set. For Mboshi, since the corpus only contains three speakers, sentences from different speakers are used in different data set respectively.
\subsection{Settings of the speech pre-trained models}

For wav2vec 2.0, the checkpoint\footnote{ https://huggingface.co/facebook/wav2vec2-base} in \cite{baevski2020wav2vec} is used. The encoder contains 7 blocks of temporal convolutions. The temporal convolutions have 512 channels with kernel sizes (10,3,3,3,3,2,2) and stride sizes (5,2,2,2,2,2,2). The context module contains 12 transformer blocks with a model dimension of 768, inner dimension of 3072 and 8 attention heads. For HuBert, the  checkpoint\footnote{https://huggingface.co/facebook/hubert-base-ls960} in \cite{hsu2021hubert} is used. The settings of its encoder and context module is the same as those for wav2vec 2.0. For the CPC model, we follow the setup in \cite{riviere2020unsupervised}. The CPC encoder is a 5-layer convolutional network with kernel sizes (10,8,4,4,4) and stride sizes (5,4,2,2,2). The CPC context module is a 1-layer GRU. It is trained on the unlabeled 960 hours LibriSpeech for 15 epochs by the Adam optimizer, with an initial learning rate of 0.0002 and a batch size of 8.

\subsection{Implementation of the frame-level AF probing tasks}
For each experiment on English or Mboshi, the following three steps were performed: 
\textbf{Step 1:} Phoneme transcriptions of TIMIT or Mboshi are mapped to articulatory features. Specifically, the ground truth classes of different articulatory features are labelled for the phonemes in TIMIT or Mboshi. For example, /\textit{i}/ is mapped to the articulatory features of \textit{\{manner:vowel, place:nil, voice:+voice, high-low:high, fr-back:front, round:-round, static:dynamic\}}; \textbf{Step2:} The speech pre-trained model is used to extract the context speech representations of the input raw audio data for the training dataset and the testing dataset. 39 dimensional MFCC speech representations are used as the baseline, with the context size of 5 windows, window size of 25ms, step size of 10ms; \textbf{Step 3:} Training the SVMs using the speech representations as input and the AF labels as target, and testing for the probing task. Multiclass classifier for each articulatory features in the probing task is implemented by \textit{sklearn's SGDClassifier} \cite{scikit-learn}. 
\subsection{Implementation of the phoneme recognition task}
For each experiment on English or Mboshi, the following two steps were performed: \textbf{Step 1:} It is same as  step 2 in section 3.3; \textbf{Step 2:} Training the phone recognition system using the speech representations as input and the phoneme transcriptions as target, and testing for the phone recognition task. The end-to-end phone recognition system is implemented by \textit{ESPnet} \cite{watanabe2018espnet}. 
\section{Results and discussion}

Tables  \ref{tab:res_timit} and \ref{tab: res_mboshi}  present the results of the frame-level AF probing tasks (upper part) and the phoneme recognition task (lower part) for TIMIT and Mboshi, respectively.
\subsection{RQ 1: Within-language scenario}

The results of the frame-level AF probing tasks on TIMIT (see Table \ref{tab:res_timit}) show that all three SSL speech pre-trained models capture more AF information than the MFCC features, with HuBert performing the best. Interestingly, the order of the best-to-worst performing speech representations is almost identical for all AFs, with HuBert outperforming the other speech representations. Compared with the averaged performance of MFCC, the performance of CPC increases 12.9\% relatively, that of wav2vec 2.0 increases 20.7\% relatively, and that of HuBert increases 34.4\% relatively. The results of the phoneme recognition tasks on TIMIT show that all three SSL speech pre-trained models achieve a lower, thus better, PER than the MFCC baseline. Compared to the baseline, the PER of CPC decreases 10.4\% relatively, that of wav2vec 2.0 decreases 46.1\% relatively, and that of HuBert decreases 59.0\% relatively. The pre-trained speech representations from HuBert outperform all other speech representations. The breakdown of the errors in substitutions, deletions, and insertions shows that HuBert has vastly fewer substitutions than the other three speech representations, which indicates that the HuBert model is able to capture phonetic information better. This result is in line with the AF probing results. Finally, the Pearson correlation between the averaged macro-averaged F1 scores of the four frame-level AF probing tasks and the four phone error rates is 0.949, showing a very strong correlation between the amount of AF information captured by the speech representations and their performance on a downstream phoneme recognition task. 
\begin{table}[t]
\centering
\caption{Macro-averaged F1 scores of the frame-level AF probing tasks and PER on the phoneme recognition tasks on TIMIT}
\label{tab:res_timit}
\resizebox{\linewidth}{!}{%
\begin{tabular}{cccccc}
\toprule
\textbf{TIMIT}       & MFCC  & CPC & wav2vec 2.0 & HuBert & \#classes \\ \midrule
voice    & 0.870 & 0.866  & 0.891          & 0.921         & 2        \\
static   & 0.669 & 0.773   & 0.786             & 0.887           & 2        \\
manner   & 0.666 & 0.733   & 0.782             & 0.842           & 7        \\
round    & 0.661 & 0.722   & 0.763             & 0.866           & 3        \\
high-low & 0.633 & 0.685   & 0.747             & 0.850           & 4        \\
fr-back  & 0.581 & 0.635   & 0.699             & 0.789           & 4        \\
place    & 0.376 & 0.621   & 0.715             & 0.840           & 6        \\ \midrule
Avg      & 0.637 & 0.719   & 0.769             & \textbf{0.856}           &          \\
Std      & 0.146 & 0.084   & 0.063             & 0.041           &          \\ \toprule
\toprule
\textbf{TIMIT}  & MFCC & CPC  & wav2vec 2.0 & HuBert \\ \midrule
\%PER          & 24.9 & 22.3 & 13.4        & \textbf{10.2}   \\
\%substitution & 14.8 & 13.3 & 7.8         & 5.5    \\
\%deletion     & 6.3  & 5.6  & 3.6         & 2.6    \\
\%insertion    & 3.8  & 3.5  & 2           & 2      \\ \bottomrule
\end{tabular}%
}
\vspace{-6mm}
\end{table}

In conclusion, speech representations extracted from speech pre-trained models, which are trained with a large amount of unlabeled raw audio data, capture more articulatory information than standard MFCCs. In addition, speech representations extracted by speech pre-trained models share similar articulatory characteristics with features extracted by MFCC. Moreover, the ability to capture articulatory feature information is strongly correlated with the accuracy of the phoneme recognition task. 

These good performances should be attributed to the deep learning architectures for sequence processing. These architectures can capture more temporal and phonetic information from the raw speech. Specifically, the context module of CPC uses GRU, which generates outputs conditionally dependent on the history information of previous outputs thus capturing information over a larger time window than MFCCs. The context modules of wav2vec 2.0 and HuBert adopt the transformer, which generates outputs that are conditionally dependent on the information of the entire input sequence, thus capturing information from a time window even larger than CPC. Meanwhile, the acoustic unit discovery (AUD) module is likely the reason for the best performance achieved by HuBert. The AUD module provided pseudo labels for HuBert during the training process, and these pseudo labels enable HuBert to effectively learn parameters which could generate speech representations with more information such as the articulatory features.
\vspace{-2mm}
\subsection{RQ 2: Cross-language scenario}
\begin{table}[t]
\centering
\caption{Macro-averaged F1 scores of the frame-level AF probing tasks and PER on the phoneme recognition tasks on Mboshi}
\label{tab: res_mboshi}
\resizebox{\linewidth}{!}{%
\begin{tabular}{cccccc}
\toprule
\textbf{Mboshi}          & MFCC  & CPC & wav2vec 2.0 & HuBert & \#classes \\ \midrule
voice    & 0.736  & 0.791  &  0.887       &  0.923     & 2        \\
fr-back  & 0.741 & 0.761   & 0.806             & 0.861           & 3        \\
round    & 0.738 & 0.766   & 0.806             & 0.861           & 3        \\
static   & 0.732 & 0.769   & 0.814             & 0.858           & 2        \\
high-low & 0.682 & 0.697   & 0.741             & 0.812           & 4        \\
place    & 0.496 & 0.545   & 0.682             & 0.786           & 5        \\
manner   & 0.466 & 0.517   & 0.598             & 0.713           & 6        \\ \midrule
Avg      & 0.656 & 0.692   & 0.762             & \textbf{0.831}          &          \\
Std      & 0.121 & 0.114   & 0.097             & 0.067           &          \\ \toprule
\toprule
\textbf{Mboshi} & MFCC & CPC  & wav2vec 2.0 & HuBert \\ \midrule
\%PER         & 56.3 & 45.9 & 32.6        & \textbf{23.0}   \\
\%substitution & 34.0   & 33.7 & 23.3        & 16.9   \\
\%deletion    & 19.6 & 9.1  & 5.0         & 3.3    \\
\%insertion   & 2.7  & 3.2  & 4.3         & 2.9    \\ \toprule
\end{tabular}%
}
\vspace{-6mm}
\end{table}
The results of the frame-level AF probing tasks on Mboshi (see Table \ref{tab: res_mboshi}) show that all three speech pre-trained models capture more AF information than MFCC, with HuBert performs the best. Compared with the averaged performance of MFCC, the performance of CPC increases 5\% relatively, that of wav2vec 2.0 increases 16.1\% relatively, and that HuBert increases 26.7\% relatively. The results of phoneme recognition tasks on Mboshi show that all three speech pre-trained models achieve a lower, thus better PER than the MFCC baseline. Compared with MFCC, the PER of CPC decreases 18.4\% relatively, wav2vec 2.0 decreases 42.1\% relatively, HuBert decreases 59.1\% relatively. Finally, the Pearson correlation in cross-language scenario is \textbf{0.990}. Thus, the amount of AF information captured by the speech representations also strongly correlated with the phoneme recognition performance in a cross-language scenario.

Although the results of these SSL speech pre-trained models on AF probing task in the cross-language scenario performed better than the baseline features, MFCCs, they are not as great as the results in the within-language scenario. 
These results indicate that indeed these SSL speech pre-trained models can be used well in cross-language scenarios. Meanwhile, these SSL speech pre-trained models also performed better than the baseline features in the phoneme recognition task.
Moreover, the PERs on Mboshi are much higher than for English.
An potential explanation of this performance gap is  the number of phonemes which is a lot smaller in TIMIT than in Mboshi, making the phoneme recognition task easier for TIMIT. Secondly, where TIMIT consists of carefully recorded read speech from multiple speakers, the Mboshi database only consists of three speakers. 
\vspace{-2mm}
\section{Conclusions}
In this work, the capability of capturing AF information of different SSL speech pre-trained models, including CPC, wav2vec2.0 and HuBert were compared and analyzed, and the correlation between the amount of AF information and the performance of the phoneme recognition have been revealed. The main conclusions are: In the within-language scenario, all speech pre-trained models capture more AF information than MFCC due to the utilization of deep learning architectures for sequence processing, and the capability of capturing AF information is strongly correlated with the phone accuracy; In the cross-language scenario, speech pre-trained models trained on English could also capture more AF information than MFCC when they extract speech representations from another language, for example, an African language. The capability of capturing AF information is also strongly correlated with the phone accuracy.

\bibliographystyle{IEEEtran}

\bibliography{template}


\end{document}